\documentclass[sigconf,nonacm]{acmart}
\usepackage[utf8]{inputenc}
\usepackage{natbib}
\usepackage{graphicx}
\usepackage{soul}
\usepackage{verbatim}
\usepackage{url}
\usepackage{amsmath}
\usepackage{algorithm,algorithmicx,algpseudocode}
\usepackage{subcaption}
\usepackage{multirow,tabularx}
\usepackage{color}
\usepackage{geometry} 
\usepackage{amssymb}
\usepackage{enumitem}
\usepackage{makecell}
\usepackage{pifont}
\usepackage{multicol}

\usepackage{filecontents}
\usepackage{listings}
\input{prolog-language}

\newcommand{\etal}{\emph{et al. }}

\renewcommand\theadfont{\normalsize}

\newcommand{\Continue}{\State \textbf{continue}}	
\algnewcommand\algorithmicforeach{\textbf{for each}}
\algdef{S}[FOR]{ForEach}[1]{\algorithmicforeach\ #1\ \algorithmicdo}
\newcommand{\xmark}{\ding{55}}  
\newcolumntype{M}[1]{>{\centering\arraybackslash}m{#1}}

\begin{document}

\title{Heuristic Approach Towards Countermeasure Selection using Attack Graphs}

\author{Orly Stan, Ron Bitton, Michal Ezrets, Moran Dadon, Yuval Elovici, Asaf Shabtai}
\affiliation{
  \institution{Dept. of Software and Information Systems Engineering \\ Ben-Gurion University of the Negev}
}

\author{Masaki Inokuchi, Yoshinobu Ohta, Tomohiko Yagyu}
\affiliation{
  \institution{Security Research Laboratories, NEC Corporation}
}

\renewcommand{\shortauthors}{O. Stan, et al.}

\begin{abstract}
    Selecting the optimal set of countermeasures is a challenging task that involves various considerations and tradeoffs such as prioritizing the risks to mitigate and costs.
    The vast majority of studies for selecting a countermeasure deployment are based on a limited risk assessment procedure that utilizes the common vulnerability scoring system (CVSS).
    Such a risk assessment procedure does not necessarily consider the prerequisites and exploitability of a specific asset, cannot distinguish insider from outsider threat actor, and does not express the consequences of exploiting a vulnerability as well as the attacker's lateral movements.
    Other studies applied a more extensive risk assessment procedure that relies on manual work and repeated assessment.
    These solutions however, do not consider the network topology and do not specify the optimal position for deploying the countermeasures, and therefore are less practical.
    In this paper we suggest a heuristic search approach for selecting the optimal countermeasure deployment under a given budget limitation.
    The proposed method expresses the risk of the system using an extended attack graph modeling, which considers the prerequisites and consequences of exploiting a vulnerability, examines the attacker's potential lateral movements, and express the physical network topology as well as vulnerabilities in network protocols.
    In addition, unlike previous studies which utilizes attack graph for countermeasure planning, the proposed methods does not require re-generating the attack graph at each stage of the procedure, which is computationally heavy, and therefore it provides a more accurate and practical countermeasure deployment planning process.
\end{abstract}


\settopmatter{printfolios=true} 

\maketitle

\section{Introduction}
In recent years we have witnessed a rapid growth in the number of vulnerabilities discovered in popular systems and applications.
When thousands of vulnerabilities revealed every month, risk management procedures must be performed automatically and continuously.

A traditional information security risk management process includes three main phases: identification, assessment and treatment. 
The identification phase begins with identifying system assets (e.g., databases, servers, data, applications, and network elements).
Next, the threats to these assets should be explored (e.g., unauthorized access, misuse of information, loss of data).
When exploring system threats, it is essential to consider both an internal and external threat actors.
The last step in this phase is monitoring the system and assets in order to identify security vulnerabilities.

In the assessment phase, the identified assets, threats and vulnerabilities are aggregated into attack paths that an attacker can take in order to compromise a system asset (i.e, a realization of one or multiple threats).
For each attack path, both the \textit{impact} to the system (i.e., the damage caused when successfully exploited) and the \textit{likelihood} (probability) of occurrence should be determined.
This probability is usually determined by considering the prerequisites and consequences of exploiting security vulnerabilities, as well as the attacker's potential lateral movements.
These two metrics define the risk for a given attack path, which is essentially the product of likelihood and impact.
The outcome of this phase is the attack surface of the system, which is represented by the different attack paths a threat actor can take to compromise system assets.
The overall risk of the system is calculated by accumulating the risk introduced by all attack paths in the attack surface.

The treatment phase involves prioritizing and implementing the appropriate security countermeasures for reducing the risk of the system.
Since eliminating all of the risk is impractical, an optimal countermeasure deployment should reduce the risk to an acceptable level with a given allocated budget. 
The outcome of this phase is usually a prioritized list of countermeasures, and the residual risk in the system after implementing those countermeasures.

Previous works for selecting the optimal countermeasure deployment can be classified into two main approaches.
The first approach utilizes the common vulnerability scoring system (CVSS)\footnote{CVSS: \url{https://www.first.org/cvss}} (e.g., \cite{viduto2012novel}),
which provides for each vulnerability a score reflecting its severity.
The score is derived from the exploitability (e.g., attack complexity, required privileges) and impact of the vulnerability (i.e., compromising confidentiality, integrity, or availability).
This approach, however, ignores the prerequisites for exploiting a vulnerability (e.g., a port that should be open in the organization firewall); does not consider the attacker lateral movement as well as the potential involvement of a vulnerability in multiple attack paths; and does not distinguish internal from external threat actors.
Therefore, although the main benefit of this approach is its efficiency (calculating the accumulated risk in a system using CVSS is very fast);
a countermeasure selection process which is based solely on CVSS cannot produce an optimal countermeasure deployment, since it is based on a limited risk assessment method.

The second approach utilizes attack graphs (e.g., \cite{chung2013nice,kotenko2016dynamic,santhanam2013identifying}).
An attack graph is a structured representation of the relationships between vulnerabilities and system assets, and between different vulnerabilities. 
Attack graphs (e.g., logical attack graph) can also represent the prerequisites and consequences of vulnerabilities, as well as the attacker's potential lateral movements (see Section~\ref{sec:ag-background} for more details). 
When generating an attack graph, the identified security vulnerabilities are aggregated into attack paths, which specify the actions a threat actor can take in order to compromise system assets.
Using attack graphs, the risk can be evaluated with respect to specific assets and threat actors.
Hence, a risk assessment method that is based on attack graph is more accurate.
However, the computational complexity for generating an attack graph is polynomial with the size of the network and the number of vulnerabilities, thus generating an attack graph is a highly time consuming task. 
Therefore, while a countermeasure selection process that is based on attack graphs can potentially provide an optimal countermeasure deployment, a naive implementation of such process on which the attack graph is re-generated at every step of the selection process is not practical.
In addition, current methods for countermeasure selection that are based on an attack graph modeling are limited because they do not consider the network topology nor vulnerabilities in network protocols.

In this paper, we suggest a heuristic method for selecting, given a budget limitation, the optimal countermeasure deployment.
The optimal countermeasure deployment is defined as the set of countermeasures, which minimizes the risk of the system within budgetary limitation.
In the proposed methods, the risk to the system is represented using an enhanced attack graph model, which in addition to software vulnerabilities considers physical network topology and vulnerabilities in network protocols.
The benefit of using this model is three-fold.
First, it provides an accurate representation of the risk to the system, which considers attacks on network protocols (such as ARP spoofing, DNS spoofing and SYN flooding). 
Second, it support advanced types of communication such as wireless communication, and thus it can be used to model cyber-attacks on networks that include IoT devices.
Third, it allow us to consider the network positions of each countermeasure.

In order to find the optimal countermeasure deployment efficiently, we model the problem as shortest path finding problem, designed an admissible heuristic to reduce the search space, and used the A* solver to find an optimal solution.
Because re-generating the attack graph at each iteration of the search process is impractical, we introduce the notion of \textit{risk equation} i.e., the risk introduced by exploiting an asset, together with possible countermeasures that can eliminate this risk; which enables us to evaluate the risk of the system under different countermeasure deployments without re-generating the attack graph.
For these reasons, the proposed method provides a more accurate and efficient countermeasure deployment planning process.

\section{\label{sec:ag-background}Attack Graphs}
An attack graph a structured representation of the vulnerabilities present in the system and their interactions with each other or system assets.
A graph-based network vulnerability analysis was first introduced by Philips and Swiler in 1998~\cite{phillips1998graph}.
In their proposed model each node in the graph represents a possible attack state (which is a combination of the current state of the host, the attacker access level, and effects of the attack so far), while edges represent an action that causes a change in the state (for example, the attacker exploits another vulnerability).
In the following decade the research on attack graphs focused mainly on enhancing the underlining models so that attack scenarios can be represented more accurately and efficiently~\cite{sheyner2002automated,jajodia2005topological,ingols2006practical,ou2006scalable} as well as enhance existing attack graph models with defenses~\cite{bistarelli2006defense,bistarelli2006strategic,roy2012attack,kordy2012attack}.

In this research we opt to use the MulVAL (a logic-based security analyzer) reasoning engine~\cite{ou2005mulval}, which is a popular, publicly available, easy to use, and extendable attack graph analysis tool.
MulVAL uses the Prolog language to represent the pre- and post conditions for exploiting a vulnerability, which allows straight forward countermeasure modeling and their automatic association to attack steps.
MulVAL\footnote{MulVAL tool: \url{http://www.arguslab.org/software/mulval.html}} models the interactions between software bugs and network configurations and automates information gathering by deploying host-based scanners. 
An example of an attack graph generated by MulVAL is presented in Figure \ref{fig:mulval-example-ag}.
MulVAL models this information using the Datalog language \cite{ou2005mulval}, which is a subset of the Prolog logic programming language.

Datalog consists of \emph{predicates} -- atomic formulas of the form:
\[
    p(t_1,...,t_k)
\]
where, each $t_i$ can be: (1) a constant (starting with a lower-case), (2) a variable (starting with an upper-case), or (3) a wild card ("\_"). 
For example, the following predicate states that the attacker (\textit{attacker}) can execute code in \textit{dbServer} under some permissions (\textit{Perm}).

\begin{equation*}
        execCode(attacker, dbServer, Perm).
\end{equation*}

Predicates are used to construct \textit{rules} that are represented using Horn clauses as follows:
\[
    P_0 :- P_1,...,P_n
\]
which are translated as: "if the predicates $P_1,...,P_n$ are true, then predicate $P_0$ is also true". 
The left part of the clause ($P_0$) is called the \textit{head} and the right part ($P_1,...,P_n$) is called the \textit{body}.
Rules with no body are called \textit{facts}. 
For example, the following rule indicates that a principal can use some host (\textit{SrcHost}) to access another (\textit{DstHost}) via network using a specific protocol (\textit{Prot}) and port (\textit{Port}, if he/she has a local access to \textit{SrcHost} and both host-based and network-based access controls allows the communication (\textit{aclH} and \textit{aclNW} predicates).
\begin{equation*}
\footnotesize
\begin{split}
netAccess&(Principal,SrcHost,DstHost,Prot,Port):- \\
		& localAccess(Principal,SrcHost,SrcUser), \\
		& aclH(SrcHost,SrcUser,SrcHost,DstHost,Prot,Port), \\
		& aclNW(SrcHost,DstHost,Prot,Port).
\end{split}
\end{equation*}

MulVAL's logical attack graph is defined as: $AG = (N_r, N_p, N_d, E, \mathcal{L}, \mathcal{G})$. 
There are three node types: derivations ($N_r$), primitive facts ($N_p$), and derived facts ($N_d$).
Derivation nodes correspond to rules and represent the logic for a fact to become true (visualized as circles); primitive facts correspond to the inputs describing the specific system (visualized as rectangles); and derived facts) are the results of applying rules on the primitive facts (visualized as diamonds).

An edge $e \in E$ can connect between a primitive or derived fact and a derivation ($e \in \{(N_p \cup N_d)\times N_r\}$), or between a derivation and a derived fact ($e \in \{N_r \times N_d\}$).

The derivation nodes imply an \textit{and} relation between their incoming nodes, which indicate the preconditions for performing the corresponding actions. The derived fact nodes, on the other hand, imply an \textit{or} relation between their incoming nodes, which represent the various ways that lead to the same consequence.

The attack graph also consists of a mapping between nodes and their labels, i.e., the predicates they represent ($\mathcal{L}$), and the node representing the attacker's goal ($\mathcal{G}$).

Attacks are simulated by querying the Datalog program, which is a collection of facts and rules describing a specific system, for the predicate representing the attacker's goal. For example, querying for $execCode(attacker, dbServer, root)$, will indicate whether the attacker can execute code with root privileges on \textit{dbServer}. 
The attack graph is constructed based on the trace generated by this query, as described by Ou \etal in \cite{ou2006scalable}.

\section{\label{sec:related-work}Related Work}
One approach for countermeasure selection is to recommend mitigation for specific events, as presented by Cuppens \etal~\cite{cuppens2006anti}. 
They suggest a decision support system to assist security administrators in selecting appropriate countermeasures, which is implemented as part of an intrusion detection system (IDS). 
The system holds a repository of previously observed attacks and countermeasures that are represented in LAMBDA (LAnguage to Model a dataBase for Detection of Attacks) -- a generic language used to describe attacks in terms of pre and post conditions. 
Countermeasures are represented in LAMBDA as follows: the pre-condition represents a system state that requires to apply the countermeasure and the post-condition which is the system state after applying the countermeasure (usually the negation of the pre-condition). 
The authors defined the concepts of \textit{correlation} and \textit{anti-correlation}. 
Two instances are correlated if the post-condition of one supports the pre-conditions of the other (i.e., these instances represent attack steps in a multistage attack scenario). 
Two instances are not-correlated if the post-condition of one contradicts the pre-conditions of the second (i.e., one instance represents a countermeasure that prevents the other instance that represents an attack).
The countermeasure recommendation process is triggered by alerts raised by the IDS. 
Upon an alert, future attack steps are identified by searching the repository for correlations, while countermeasures are identified by searching for anti-correlations.

In this work, we use LAMBDA for the representation of mitigation actions. 
The structure of LAMBDA (i.e., the pre and post conditions) is very similar to MulVAL's attack graph modeling, which also represents attacks according to pre and post conditions (see Section \ref{sec:ag-background}). 
Thus, enables easy representation and mitigation actions to attack graph nodes matching (see Subsection \ref{subsec:cm-id}).

Other approaches suggest selecting a countermeasure plan to reduce the global risk in a system, and not just to mitigate a specific event.
Viduto~\etal~\cite{viduto2012novel} presented an approach for recommending efficient countermeasure set in terms of both cost and risk. 
Their approach is divided into two parts: (1) risk assessment process that results in an initial countermeasure set, and (2) an optimization process that refines the selected set using a multi-objective optimization algorithm (Tabu-search).
The risk assessment process consists of 11 steps, during which: the organization's assets are identified and scanned for vulnerabilities and existing exploits; vulnerabilities are mapped to their impact based on their CVSS scores and are matched to threats identified in the organization; a likelihood table is computed to indicate the probability for a threat to materialize given that the attacker exploited a vulnerability; finally, the total risk in the system is computed and an initial countermeasure set ids selected from a predefined generic list.
In the second phase of their approach, the initial countermeasure set is used as the initial state to the heuristic multi-objective Tabu search algorithm, which finds a countermeasure set that minimizes both the countermeasures implementation cost and the risk in the system.
Although their approach is designed to consider the two major aspects in countermeasure selection, the mitigated risk mitigated and and cost, the risk assessment process is mostly manual and requires experts' knowledge.

Unlike Viduto \etal, we follow previous works that suggested automating the risk assessment phase by using attack graphs.

Kijsanayothin and Hewett~\cite{kijsanayothin2010analytical} presented an analytical approach for using attack graphs for network protection. 
Their approach chooses the most preferred countermeasures to eliminate attack paths using CP-Net (a graphical model for representing qualitative conditional preference relationship among decision variables).
However, it cannot provide a recommendation that satisfies a given budget. 
Moreover, they don't consider the amount of risk each countermeasure mitigates.

Santhanam~\etal~\cite{santhanam2013identifying} used preference graph to generate the most preferred and effective (in terms of attack paths elimination) countermeasure strategy. 
They introduced the C$^3$I-net (credibility-constrained conditional importance network), which is based on the CI-net (conditional importance network) -- a language for representing ordinal preferences over a set of goods. In their approach, the C$^3$I-net is converted to a preference graph, which enables to iteratively discover the most preferred strategies. The effectiveness of each strategy is evaluated by generating an attack graph according to its effect. If the attacker's goals are satisfied, the strategy is considered effective. The optimal solution is the first discovered strategy that is also effective.

Chung~\etal~\cite{chung2013nice} presented the network intrusion detection and countermeasure selection (NICE) framework. 
NICE is a multi-phase distributed vulnerability detection, measurement, and countermeasure selection mechanism aimed at preventing virtual machines (VMs) in a cloud from being compromised. 
The authors used a variation of MulVAL for generating the attack graph that is able to represent IDS alerts. 
The attack graph is generated based on vulnerabilities discovered in offline or online scans and the alerts raised, and is regenerated upon the discovery of new vulnerabilities or countermeasure deployment.
Similar to \cite{cuppens2006anti} \etal, countermeasures are selected per alert from a predefined countermeasures pool. 
However, in this work, the probability of vulnerability exploitation (based on its CVSS base metric) is taken into consideration.
Upon an alert, the probabilities of the corresponding node and the rest of the nodes in the path to the goal are updated.
Based on these probabilities, the benefit of deploying each suitable countermeasure is calculated. Finally the optimal countermeasure is selected based on the return on investment (ROI) metric.

Gonzalez-Granadillo~\etal~\cite{gonzalez2017attack} used the return on response investment (RORI) metric, which indicates the gain earned by implementing a countermeasure (i.e., response) relative to its cost, and introduced its stateful version (StRORI) that includes consideration in the system's state at the time of evaluation.
The authors present a mitigation model to handle ongoing and potential attack, which has two operation modes. 
The first is the \textit{preventive mode}, during which an initial set of countermeasures is created based on attack graph nodes coverage.
The second is the \textit{reactive mode}, during which the risk associated with each attack graph node is updated based on reports of real attacks identified in the system, and new countermeasures are selected (based on the StRORI metric) and enforced. 

Kotenko and Doynikova \cite{kotenko2016dynamic} defined a method for countermeasure selection that utilized attack graph-based metrics.
The attack graph used by their method is \cite{jajodia2005topological}. 
Their method is based on many open standards (e.g., CVE, CVSS) and relies on existing countermeasures, such as firewalls and IDS, to provide information regarding security events.
Similar to \cite{cuppens2006anti,chung2013nice}, the recommendations are event-driven (i.e., the countermeasure selection process is triggered by a security event).
An attack graph is generate, based on the various information sources, to represent the system's security state. 
Upon an event, attacker and attack related metric, (such as: attack probability, attacker skills, and attack impact) are calculated based on the attack graph. Then, a \textit{countermeasure index} is computed for each countermeasure, which reflects its benefit in light of the expected losses, and the countermeasure with the maximal index value is selected. To measure the countermeasure effectiveness, the attack probability metric is recalculated, which requires to modify the attack graph accordingly.

Table~\ref{tab:pre-work} summarizes the similarities and differences of the above mentioned methods and our approach. 
We examine the works based on their definition of countermeasures (generic, e.g., use firewall of some brand, or specific actions, e.g., update a specific rule in a specific firewall instance), countermeasure-vulnerability matching method, and countermeasure selection considerations (minimizing the risk in the system, minimizing deployment cost, operator preferences, risk introduced by the countermeasure, and in which position should the countermeasure be placed). 
Each of these works addressed one or more major considerations, however non of them provided an holistic solution to provide a detailed countermeasure plan that meets the organization's requirements. 
Moreover, most of the works manually matched countermeasure to vulnerabilities, or didn't provide any details regarding their method.

\begin{table*}
    \centering
    \caption{Comparison with previous works.}
    \scriptsize
    \renewcommand{\arraystretch}{1.1}
    \begin{tabular}{|>{\centering}m{0.15\textwidth}|>{\centering}m{0.12\textwidth}|>{\centering}m{0.15\textwidth}|c|>{\centering}m{0.05\textwidth}|c|c|c|c|c|}
         \hline
         \multirow{2}{=}{\centering Work} & \multirow{2}{=}{\centering CM Definition} & \multirow{2}{=}{\centering CM Identification} & \multirow{2}{*}{AG} & \multirow{2}{=}{\centering AG re-gen.} & \multicolumn{5}{c|}{Considerations}  \\
         \cline{6-10}
         & & & & & Risk Min. & Cost & Pref. & CM Risk & Pos.\\
         \hline
         Cuppen \etal \cite{cuppens2006anti} & Depends on the LAMBDA spec. & Search for anti-correlation in the IDS alerts DB & & & &  &  &  & \checkmark \\
         \hline
         Viduto \etal \cite{viduto2012novel} & Generic & Predefined mapping & & & \checkmark & \checkmark &  & \checkmark &  \\
         \hline
         Santhanam \etal \cite{santhanam2013identifying} & Mitigation action & -- & \checkmark & \checkmark & &  & \checkmark &  &  \\
         \hline
         Gonzalez-Granadillo \etal \cite{gonzalez2017attack} & -- & -- & \checkmark & \xmark & \checkmark & \checkmark &  & &  \\
         \hline
         Kijsanayothin and Hewett \cite{kijsanayothin2010analytical} & Mitigation action & -- & \checkmark & \xmark & & \checkmark & \checkmark &  &  \\
         \hline
         Chung \etal \cite{chung2013nice} & Generic & -- & \checkmark & \checkmark & \checkmark & \checkmark &  & \checkmark & \checkmark \\
         \hline
         Kotenko and Doynikova \cite{kotenko2016dynamic} & Mitigation actions & Predefined mapping & \checkmark & \checkmark & & \checkmark & & \checkmark &  \\
         \hline \hline
         Our approach & Mitigation action / CM & Automatic (Prolog reasoning) & \checkmark & \xmark & \checkmark & \checkmark &  &  & \checkmark \\
         \hline
         \multicolumn{3}{l}{CM = Countermeasure} &
         \multicolumn{7}{l}{AG -- use attack graph for risk assessment} \\
         \multicolumn{3}{l}{AG re-gen -- whether the attack graph is regenerated each time the risk is assessed} &
         \multicolumn{7}{l}{Risk Min. -- minimizes the risk in the system} \\
         \multicolumn{3}{l}{Cost -- minimizes the cost of the plan} &
         \multicolumn{7}{l}{Pref. -- select countermeasures with regard to the operator preferences} \\
         \multicolumn{3}{l}{CM Risk -- consider the risk introduces by countermeasures} &
         \multicolumn{7}{l}{Pos. -- consider the deployment position of the countermeasure in the system}
    \end{tabular}
    \label{tab:pre-work}
\end{table*}

We aim to provide a method for constructing a countermeasure plan that: (1) is specific as possible (i.e., specifies where to deploy/perform each countermeasure/mitigation action), (2) meets the organization's budget, (3) corresponds as much as possible to the operator's preferences, and (4) minimizes as much as possible the risk in the system. 
The risk introduced by countermeasures is not considered within the scope of this work.
In addition, unlike previous works, we distinguish between mitigation actions and countermeasures and consider them both. 
This enables an efficient countermeasure-vulnerability matching as will be described in the following sections.

\section{\label{sec:mitigation-action}Mitigation Action and Countermeasure Modeling}
\subsection{Overview}
In this work, we define countermeasures as specific security products (e.g., Palo Alto Firewall) that are mainly characterized by two attributes: the defense mechanism they implement and the set of mitigation actions they provide. 
A defense mechanism defines a family of security products (e.g., Firewall, antivirus). 
Mitigation action, on the other hand, is a specific functionality that is provided by the countermeasure and is not associated with a specific security product (e.g., update a Firewall rule to disable/enable specific communication).

Note that a security breach can be eliminated by different mitigation actions.
For example, code execution due to vulnerability in a local program could be eliminated by both patching the vulnerable program or installing an anti-virus to prevent it. 
The categorization of countermeasures in both general (defence mechanism) and specific (mitigation action) attributes allows more efficient matching process between security issues and countermeasure. 

The following sections provide detailed description of the countermeasures and mitigation actions modeling. 
Moreover, Section~\ref{subsec:ma} presents a methodology for defining specific mitigation actions based on MulVAL attack graph modeling.

\subsection{\label{subsec:ma}Mitigation action}
Our modeling of mitigation actions is inspired by LAMBDA~\cite{cuppens2006anti}, and consists of the following fields:

\begin{itemize}
    \item ID -- unique identifier.
    \item Countermeasure type: the defense mechanism corresponding to the mitigation action.
    \item Action -- description of the mitigation action.
    \item Post-conditions -- the cancelled conditions and side-effects of applying the action.
    \item Pre-conditions -- the conditions that must be met in order to apply the action.
    \item Position -- pointer to the location of the new security product in the network.
\end{itemize}

The pre- and post-conditions are represented by the same Prolog predicates that are used to generate the attack graph. 
The primary post-condition represents the fact that is cancelled by the mitigation action (thus it is written in negation). 
Other post-conditions are optional and represent the ``side-effects" of performing the mitigation action. 
The pre-conditions represent the conditions that must hold in order to perform the mitigation action.
We chose to represent the mitigation actions' pre- and post-conditions using Prolog predicates in order to be able to automatically match them with each attack graph node, and also to use the corresponding query traces to infer the network position that the mitigation action should be applied.
Examples for specific mitigation actions representation is presented in Appendix~\ref{app:ma-def-example}.
Following is a description of a methodology for modeling mitigation actions based on a given Prolog program (i.e., attack graph modeling).

\subsubsection{Modeling Methodology}
\begin{figure}
\includegraphics[scale=0.30]{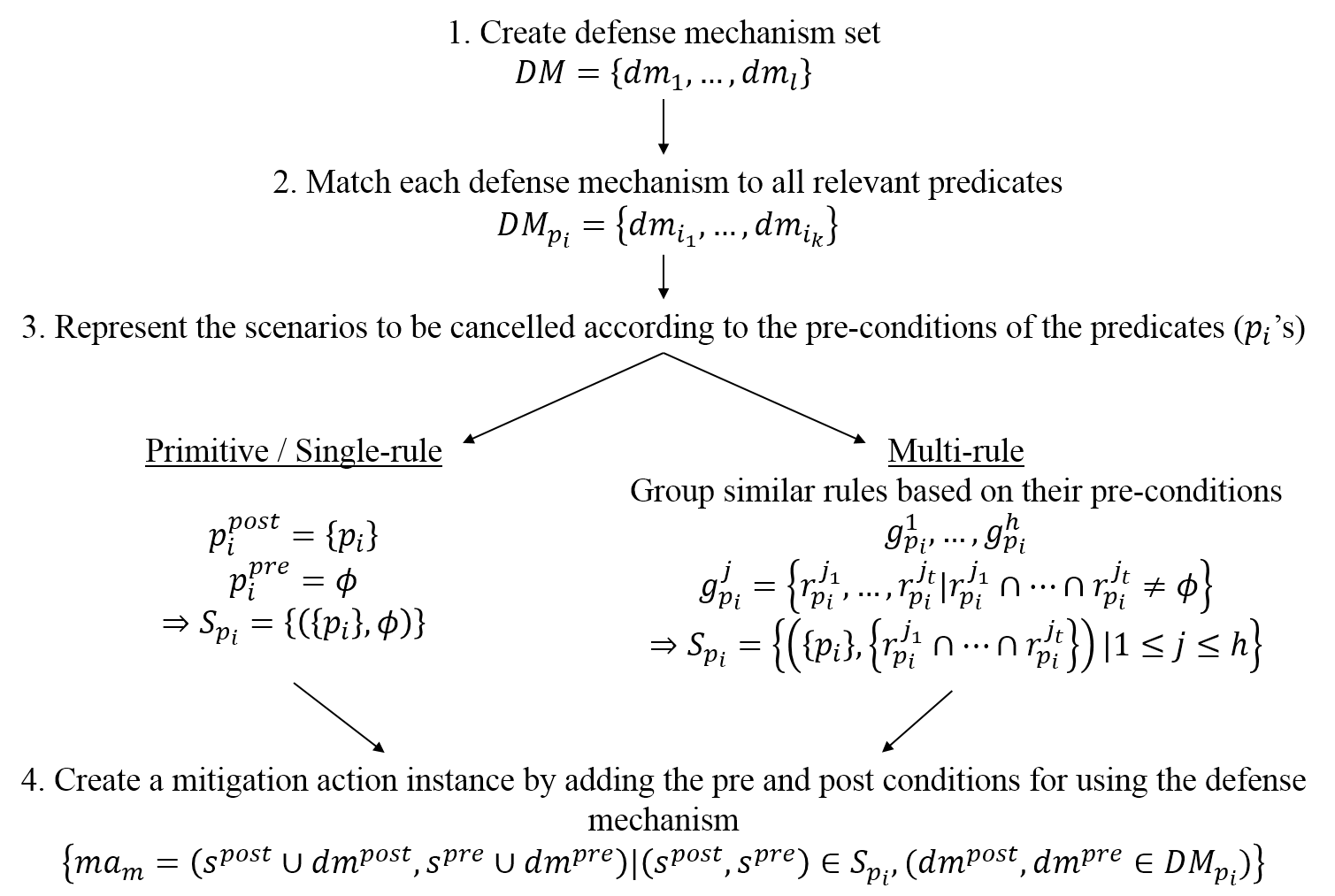}
	\caption{Mitigation action definition methodology.}
    \label{fig:ma-methodology}
\end{figure}

The process of defining mitigation actions is illustrated in Figure~\ref{fig:ma-methodology}.
First, defense mechanisms are enumerated in order to create a general set of remedies (denoted by $DM$), which are then associated with all of the predicates they can eliminate.
Note that $DM_{p_i}$ is the set of defense mechanisms that can eliminate the fact represented by predicate $p_i$.
In this work, we consider the following defense mechanisms: firewall (host and network based), software patch, antivirus software, and intrusion prevention system (host and network based).

Next, the conditions that enable the existence of the facts represented by the predicate $p_i$ are identified. 
We distinguish between two cases: (1) primitive predicates or predicates that are derived by only one rule (referred to as single-rule predicates) and (2) predicates that are derived by multiple rules (referred to as multi-rule predicates).
Predicates of the first case represent only one possible scenario, thus we consider only the pre-conditions of the defense mechanism.
For example, \textit{vulHost} is a primitive predicate that indicates the existence of a specific vulnerability in a specific program running in some host:
\[
    vulHost(Host, VulID, Program, Range, Consequence).
\]
A vulnerability can be handled by two defense mechanisms from our set: software patch or network-based IPS. 
Each of them requires that there exists a patch/IPS rule that can identify the specific vulnerability (denoted by \textit{VulID}).

Predicates of the second case might represent more than one possible scenario that leads to the same consequence. 
In this case, all of the rules are first grouped by their pre-conditions such that the rules in each group (denoted by $g^{p_i}_j$) represent similar scenarios. 
Each group is a candidate for a mitigation action, and the set of common pre-conditions is part of the corresponding mitigation action's pre-conditions. 
For example, \textit{aclNW} is a predicate that describes network communication:
\[ aclNW(SrcHostOrSubnet, DstHostOrSubnet, Protocol, Port). \]
This predicate describes four different types of communications: (1) between two hosts (inside the same subnet or between subnets); (2) between two subnets; (3) between a  specific host and an entire subnet; and (4) between an entire subnet and a specific host. 
Thus, the rules deriving the \textit{aclNW} are divided into four groups, which can be cancelled by a network-based firewall. 
Note that the post-condition for each predicate $p_i$ is the predicate itself (i.e., the fact that is cancelled by the mitigation action).

Finally, mitigation action instances are created from the combination of the previously extracted scenarios and the defense mechanisms that can eliminate them. 
Note that for defense mechanism that requires installation (e.g., firewall or IPS) two mitigation actions are created -- for the case that the defense mechanism should be installed (in this case another pre-condition is added to indicate that it is not installed yet) and for the case that it is installed and needs to be updated (in this case another pre-condition is added to indicate that the defence mechanism exists).
A detailed example of defining mitigation actions for the \textit{vulHost} and \textit{aclNW} predicates is given in Appendix~\ref{app:ma-def-example}.

Mitigation actions also contain a specification of how to extract the network position where to deploy the defense mechanism from the corresponding query trace. 
The deployment position can be one of the following: (1) the vulnerable host (e.g., installing an HIPS or patching a program; (2) an existing security product (e.g., updating a firewall rule or an NIPS); and (3) a network position (e.g., when installing a new firewall to separate two existing subnets).

\subsection{Countermeasures}
Countermeasures are characterized by more technical details regarding security products such as: cost, manufacturer, and are associated to the set of mitigation actions they support.
Each product can support one or more mitigation actions and consists of the following fields:

\begin{itemize}
    \item ID -- unique identifier.
    \item Manufacturer -- the name or identifier of the manufacturing company (e.g., Cisco).
    \item Product ID -- the name or identifier of the specific product of the manufacturing company (e.g., Snort).
    \item Deployment Cost -- the cost of the product and the deployment expenses.
    \item Coin -- the currency of the deployment cost value (e.g., USD).
    \item Mitigation Action IDs -- list of mitigation action id's supported by the product.
\end{itemize}

The modeling of new countermeasures is based on market research. 
New countermeasures are added if new products appear on the market.
The products modeled must provide one or more mitigation actions. 
Note that new types of security products can also lead to the creation of new mitigation actions. 
In light of our modeling method, by modeling mitigation actions and countermeasures from a group of existing security product types, we prevent the creation of imbalance between the two.

\section{\label{sec:risk-ass}Risk Assessment}
Before executing the search to find a cost-effective countermeasure plan, a risk assessment process should be conducted in order to estimate the current state of the system and understand what are the options to enhance it. 
We chose to perform an attack graph-based risk assessment approach; more specifically we chose to use MulVAL~\cite{ou2005mulval} with our extended network modeling\footnote{Reference removed not to break the anonymity.}.

The following subsections describe how relevant countermeasures are identified based on the generated attack graph, and and alternative representation of the attack graph, which also incorporate the relevant countermeasures, that allows more efficient calculation of the system's risk under the deployment of a specific countermeasure plan.

\subsection{\label{subsec:cm-id}Identifying Relevant Countermeasures}
Countermeasures identification is a two steps process that is based on matching mitigation actions to the attack graph nodes. 
The mitigation action matching process is also based on Prolog reasoning, similar to the attack graph generation~\cite{ou2005mulval}. 
Note however, that the Prolog program used for this task is slightly different than the one used to construct the attack graph.  

The Prolog program for mitigation action matching consists of the input used to generate the attack graph (i.e., facts about the system) and a set of rules that represent the various mitigation actions. 
Each mitigation action (as defined in Section~\ref{sec:mitigation-action}) is converted into a rule in the following format:
\begin{equation}
    Postcondition :- Precondition_1, ... , Precondition_n
\end{equation}

where $Postcondition$ corresponds to the predicate that the mitigation action eliminates, with an additional argument to indicate the mitigation action ID:
\begin{equation}
    \label{eq:ma-postcond}
    p(t_1, ..., t_n) \Rightarrow p(MAID, t_1, ..., t_n)
\end{equation}

The preconditions on the other hand, remain the same as in the mitigation action description.
For example, consider the following predicate, which represents that a communication (using protocol \textit{Prot} on port \textit{Port}) between two subnets (\textit{SrcSubnet} and \textit{DstSubnet}) is enabled:

\begin{equation}
    aclNW(SrcSubnet, DstSubnet, Prot, Port)
\end{equation}

A possible mitigation action is updating a firewall rule and disable this communication. 
The only precondition for applying this mitigation action is the existence of a firewall between those two subnets (represented by an \textit{isFirewall} predicate); thus, the rule representing this mitigation action is as follows (assuming this mitigation action's ID is 1):

\begin{equation}
    \label{eq:nb-fw-ma}
    \begin{split}
        aclNW&(1,SrcSubnet,DstSubnet,Prot,Port) :- \\
        & isFirewall(FW,SrcSubnet,DstSubnet).
    \end{split}
\end{equation}

Given an attack graph node, a query is sent to the Prolog program based on the node's description (according to Equation \ref{eq:ma-postcond}). The results of the query will contain all of the mitigation action IDs that can be applied to eliminate this node.

The next step is to understand where to perform the mitigation action in the network.
The network position is extracted based on the \textit{position} field of each mitigation action (see Section~\ref{sec:mitigation-action}) that instructs how to find the relevant arguments from the Prolog trace.
For example, the position of the mitigation action represented by Equation \ref{eq:nb-fw-ma} is the value of the \textit{FW} variable in the \textit{isFirewall} predicate.

Finally, after obtaining the relevant mitigation actions IDs, all of the corresponding countermeasures are retrieved from the countermeasures repository and construct the initial node of the search. 
The network position of a countermeasure is similar to the position of its corresponding mitigation action.

\subsection{\label{subsec:risk-eq}Risk Calculation}
The risk in the system is computed using \textit{risk equations} that are derived from the attack graph during the pre-processing phase to the search. 
The risk is measured by the ease of taking each attack path.
Risk equation is a mathematical representation that incorporates the risk introduced by the attacker's actions towards his/her goal and the possible countermeasures that can thwart these action. 

\begin{figure*}
	\begin{subfigure}[h]{0.35\textwidth}
	    \includegraphics[scale=0.35]{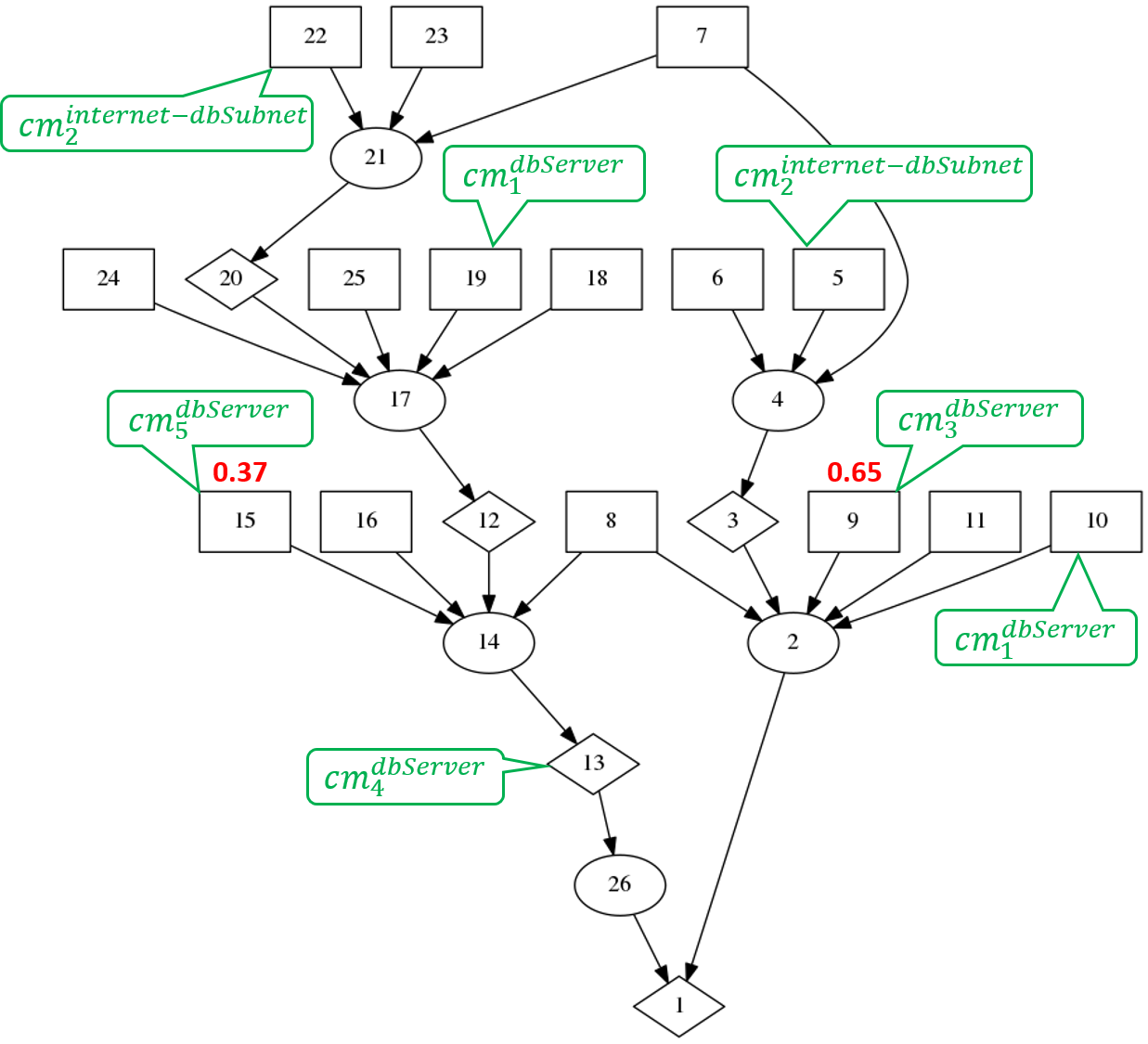}
	\end{subfigure}
	\hspace{30pt}
	\begin{subfigure}[h]{0.4\textwidth}
		\tiny
		\begin{tabular}{|c|m{0.65\textwidth}|}\hline
		    Node & Description \\ \hline
    		1 & dos(attacker, dbServer)  \\ \hline
            2 & Rule 1: DoS by remote exploit  \\ \hline
            3 & netAccess(attacker, attackerHost, dbServer, tcp, 1521)  \\ \hline
            4 & Rule 2: Principal can access a host from a neighbor  \\ \hline
            5 & aclNW(attackerHost, dbServer, tcp, 1521) 
            \\ \hline
            6 & aclH(attackerHost, admin, attackerHost, dbServer, tcp, 1521)  \\ \hline
            7 & localAccess(attacker, attackerHost, admin)  \\ \hline
            8 & malicious(attacker)  \\ \hline
            9 & vulHost(dbServer, 'CVE-2019-2510', oracle\_mysql, remoteExploit, dos) 
            \\ \hline
            10 & aclH(dbServer, admin, attackerHost, dbServer, tcp, 1521) 
            \\ \hline
            11 & networkService(dbServer, oracle\_mysql, tcp, 1521, admin) \\ \hline
            12 & localAccess(attacker, dbServer, admin) \\ \hline
            13 & execCode(attacker, dbServer, admin) 
            \\ \hline
            14 & Rule 3: Code execution by local exploit \\ \hline
            15 & vulHost(dbServer, 'CVE-2017-8714', windows\_server\_2012, localExploit, completePrivEsc)  
            \\ \hline
            16 & localService(dbServer, windows\_server\_2012, admin) \\ \hline
            17 &  Rule 4: Principal has account and can access host via network  \\ \hline
            18 & isLoginService(remote\_desktop)  \\ \hline
            19 & aclH(dbServer, admin, attackerHost, dbServer, rdp, 3389) 
            \\ \hline
            20 & netAccess(attacker, attackerHost, dbServer, rdp, 3389)  \\ \hline
            21 & Rule 2: Principal can access a host from a neighbor  \\ \hline
            22 & aclNW(attackerHost, dbServer, rdp, 3389) 
            \\ \hline
            23 & aclH(attackerHost, admin, attackerHost, dbServer, rdp, 3389)  \\ \hline
            24 & networkService(dbServer, remote\_desktop, rdp, 3389, admin)  \\ \hline
            25 & hasAccount(attacker, dbServer, admin) \\ \hline
            26 & Rule 5: DoS by code execution \\ \hline
		\end{tabular}
	\end{subfigure}
	\caption{MulVAL example: code execution attack graph.}
    \label{fig:mulval-example-ag}
\end{figure*}

For example, consider the attack graph in Figure~\ref{fig:mulval-example-ag}, which illustrates a scenario of a remote attacker that exploits the lack of access control between the Internet and a vulnerable database server in order to crash it. 
Two vulnerabilities present in the database server: CVE-2017-8714 with risk score of 0.37 and CVE-2019-2510 with risk score of 0.65. 
We consider four types of mitigation actions: (1) update host-based firewall; (2) install network-based firewall; (3) patch a program; and (4) install Antivirus software. 
Table~\ref{tab:example-cm} specifies the position and notation of these countermeasures.

\begin{table}[t]
        \centering
        \caption{Examples of countermeasures.}
        \scriptsize
        \begin{tabular}{|m{0.12\textwidth}|m{0.09\textwidth}|m{0.07\textwidth}|c|c|}
            \hline
             Countermeasure & Mitigation action & Position & Notation & Cost(\$)\\
             \hline
             Host-based firewall & update rule & $dbServer$ & $C_1$ & 20 \\
             \hline
             Network-based firewall & install new component & $internet$-$dbServer$ & $C_2$ & 500 \\ 
             \hline
             \multirow{2}{=}{Patch} & \multirow{2}{=}{Patch a program} & \multirow{2}{=}{$dbServer$} & $C_3$ & \multirow{2}{*}{10} \\
             \cline{4-4}
             & & & $C_5$ & \\
             \hline
             Antivirus & Install Antivirus software & $dbServer$ & $C_4$ & 50 \\
             \hline
        \end{tabular}
        \label{tab:example-cm}
    \end{table}

Algorithm \ref{alg:build-risk-eq} describes the process of building the risk equations for a given attack graph (denoted by $AG$), that builds the equations for each node in the attack graph. It maintains two sets: $processed$ that keeps all of the nodes that were already processed (i.e, their equation was constructed), and $unprocessed$ that keeps all of the remaining nodes. In each iteration a processable node (i.e., a node that all of its parents are processed) is chosen and its equation is constructed, until all nodes are processed.

\begin{algorithm}
	\caption{Build Risk Equations.}
	\scriptsize
	\label{alg:build-risk-eq}
	\begin{algorithmic}[1]
        \State $processed \gets \{Risk(n) | n \in AG_{LEAF}\}$
        \State $unprocessed \gets AG \setminus AG_{LEAF}$
        \While{$!unprocessed.isEmpty()$}
            \While{$!unprocessed.hasProcessableNode()$}
                \State $n \gets unprocessed.pop()$
                \State $n.updateMA()$
                \If{$\forall k \in n_{in}, k \in processed$}
                    \State $processed \gets processed \cup \{Risk(n)\}$
                \Else
                    \State $unprocessed.push(n)$
                \EndIf
            \EndWhile
            \State $removeCycles(AG)$
        \EndWhile
	\end{algorithmic}
\end{algorithm}

Note that if the attack graph contains cycles, the iteration will never finish. 
Cycles are caused by OR nodes that are derived by some interaction rule and also lead to a pre-condition for the same rule.
To break these cycles, the algorithm identifies unprocessed OR nodes with more than one parent that at least one of them is processed, and removes the edges between the unprocessed AND parents to the unprocessed OR node.
\begin{equation}
\scriptsize
    Risk(LEAF) = \left( \prod_{m \in LEAF_{CM}} m \right) \cdot P(LEAF)
    \label{eq:leaf-prob}
\end{equation}

\begin{equation}
\scriptsize
    Risk(AND) = \prod_{n \in AND_{in}} Risk(n)
    \label{eq:and-prob}
\end{equation}
\begin{equation}
\scriptsize
    \begin{split}
    Risk&(OR) =  \left( \prod_{m \in OR_{CM}} m \right) \cdot \left( \sum_{i=1}^{R} Risk(OR_{in}[i]) - \right. \\
    & \left. \sum_{i_1<i_2} Risk(OR_{in}[i_1]) \cdot  Risk(OR_{in}[i_2]) + \cdots + \right. \\
    & \left. (-1)^{R} \cdot \sum_{i_1<\cdots<i_{R-1}} Risk(OR_{in}[i_1])\cdot ... \cdot Risk(OR_{in}[i_{R-1}]) +  \right. \\ 
    & \left. (-1)^{R+1}\cdot \prod_{i=1}^{R} Risk(OR_{in}[i]) \right) \\
    & \\
    & where, ~ R = |OR_{in}|
    \end{split}
    \label{eq:or-prob}
\end{equation}

\begin{equation}
\scriptsize
    m = 
    \begin{cases}
        0, & \text{if countermeasure m is implemented} \\
        1, & \text{if countermeasure m is \textbf{not} implemented}
    \end{cases}
    \label{eq:m-value}
\end{equation}

The implementation of the $Risk$ function is given in Equations \ref{eq:leaf-prob}, \ref{eq:and-prob}, \ref{eq:or-prob}, and \ref{eq:m-value}. 
If the given attack graph node represents a vulnerability, then its risk score is calculated according to the vulnerability's CVSS score. 
Otherwise, the node's risk score is 1 (we assume than any other fact can be materialized in any situation).

The risk score of a vulnerability node represents its probability to be exploited, and is calculated according to the following CVSS metrics, which provides information about the pre-conditions of vulnerability exploitation that is not represented in the attack graph:

\begin{equation}
\scriptsize
    Risk(n) = 
    \begin{cases}
        AC \cdot UI,& \text{if CVSS v.3} \\
        AC \cdot Au, & \text{if CVSS v.2}
    \end{cases}
\end{equation}

CVSS provides both qualitative and quantitative values for the metrics it defines, these values are fixed and are given in the CVSS specification. 
The values of the above mentioned exploitability metrics are in the range [0,1], thus $Risk(n)\in [0,1]$. 
This enables us to consider this score as a probability in future calculations.

Equation \ref{eq:risk-eq-example} presents the calculation of the risk equation for the goal node (node 1) in the attack graph from Figure \ref{fig:mulval-example-ag}.

\begin{equation}
    \scriptsize
        \label{eq:risk-eq-example}
        \begin{split}
            R&isk(n_1) = Risk(n_{26}) + Risk(n_2)\\
                & = (C_4 \cdot Risk(n_{13})) + (Risk(n_8) \cdot Risk(n_3) \cdot Risk(n_9) \cdot Risk(n_{11}) \cdot Risk(n_{10})) \\
                & = (C_4 \cdot Risk(n_{14})) + (Risk(n_4) \cdot (0.65 \cdot C_3) \cdot C_1) \\
                & = (C_4 \cdot (Risk(n_{15}) \cdot Risk(n_{16}) \cdot Risk(n_{12}) \cdot Risk(n_8))) + (0.65 \cdot C_3 \cdot C_1 \cdot \\
                & \qquad \cdot (Risk(n_6) \cdot Risk(n_5) \cdot Risk(n_7))) \\
                & = (C_4 \cdot ((0.37 \cdot C_5) \cdot Risk(n_{17}))) + (0.65 \cdot C_3 \cdot C_1 \cdot (C_2)) \\
                & = (C_4 \cdot (0.13 \cdot C_5 \cdot (Risk(n_{24}) \cdot Risk(n_{20}) \cdot Risk(n_{25}) \cdot Risk(n_{19}) \cdot Risk(n_{18})))) + \\
                & \qquad +(0.65 \cdot C_3 \cdot C_1 \cdot C_2) \\
                & = (C_4 \cdot (0.37 \cdot C_5 \cdot (Risk(n_{21}) \cdot C_1))) + (0.65 \cdot C_3 \cdot C_1 \cdot C_2) \\
                & = (C_4 \cdot (0.37 \cdot C_5 \cdot (C_1 \cdot (Risk(n_{22}) \cdot Risk(n_{23}) \cdot Risk(n_7))))) + (0.65 \cdot C_3 \cdot C_1 \cdot C_2) \\
                & = (C_4 \cdot (0.37 \cdot C_5 \cdot (C_1 \cdot (C_2)))) + (0.65 \cdot C_3 \cdot C_1 \cdot C_2) \\
                & = (0.37 \cdot C_4 \cdot C_5 \cdot C_1 \cdot C_2) + (0.65 \cdot C_3 \cdot C_1 \cdot C_2)
        \end{split}
    \end{equation}

\section{\label{sec:heuristic}Heuristic Algorithm}
In this section, we present a heuristic approach for finding a countermeasure plan for a specific system under a given budget limitation that minimizes the risk in the system as much as possible. 
This approach is based on the attack graph-based risk assessment products described in~\ref{subsec:risk-eq}.
Following is a formal definitions of the problem and a description of a heuristic solver that solves it.

\subsection{\label{subsec:problem-def}Problem Definition}
We suggest to model the countermeasure selection problem as a graph path finding problem and use a heuristic solver to find an optimal solution as defined by Pohl in \cite{pohl1970heuristic}. 
Given a set of countermeasures ($CM$) and a budget limitation ($Budget$), we define the path finding problem as follows:

\begin{itemize}
    \item $X = \{x_1,x_2, ... ,x_n\}$ is the set of nodes (also referred to as \textit{states}), where $x_i = \{cm_{ID}^{pos} | cm_{ID} \in CM\}$ for $1 \leq i \leq n$; $pos$ corresponds to the network position where the countermeasure should be deployed. 
    \item $E = \{(x_i,x_j)|x_i,x_j \in X, x_j \in \Gamma(x_i)\}$ is the set of edges between the nodes in $X$, where $\Gamma$ is the mapping between a node to its successor.
    \item $\Gamma(x) = \{x_1,x_2, ... , x_{|x_i|} |\forall k, 1 \leq k \leq |x_i|: x_{i_k}=x \setminus \{x[k]\}\}$ is the successor mapping (also referred to as \textit{expand function}). The successors of a specific node are created by omitting one countermeasure each time.
    \item The initial node contains all of the available and relevant countermeasures to reduce the risk in the system.
    \item A terminal node (also referred to as \textit{goal node}) is a node $x$ such that $Cost(x)=\sum_{i=1}^{|x|}Cost(x[i]) \leq Budget$; i.e., the cost of deploying the countermeasure plan it represents stands within the budget limitation.
    \item $f(x) = g(x) + h(x)$ reflects the estimated risk to the system under the deployment of a subset of the countermeasures in $x$ (also referred to as \textit{f-value}); where $g(x)$ is the risk to the system under the deployment of all countermeasures in $x$ (also referred to as \textit{g-value}), and $h(x)$ is an estimation of the risk added by removing countermeasures from $x$ in order to comply with the given budget limitation (also referred to as \textit{h-value}).
\end{itemize}

\subsection{\label{subsec:astar}Heuristic Solver}
To solve the path finding problem defined above and find an optimal countermeasure plan, we used the A* algorithm. 
A* is a best-first search algorithm that aims to find a path from a given initial node to a goal node with the smallest possible cost (i.e., the sum of the path's edges cost is the smallest). 
In our context, the sum of a path corresponds to the risk in the system under the deployment of the plan represented by the goal node; i.e., each edge corresponds to the new risk introduced to the system by omitting a countermeasure.

To perform the search, the algorithm maintains two sets: open list and closed list. 
The closed list contains all of the nodes that were already selected to participate in a path, while the open list holds all of the potential nodes to continue a path.
In each iteration, the algorithm selects the best node from the open list and adds it to its corresponding path. 
It also expands the open list with all of the candidate nodes to continue this path (i.e., the result of applying the expand function on the selected node).
The open list is implemented as a priority queue that sorts the nodes by their f-value.
Algorithm~\ref{alg:astar_solver} presents the A* solver for the countermeasure selection problem.

\begin{algorithm}[t]
\caption{A* Solver}\label{alg:astar_solver}
\scriptsize
    \begin{algorithmic}[1]
        \State\textbf{Inputs:}
            \State $\qquad$ $RiskEq \gets RiskEquations$
            \State $\qquad$ $initial \gets RelevantCountermeasures$
            \State $\qquad$ $\text{B} \gets \text{Budget}$
        \State\textbf{Initialize:}
            \State $\qquad$ $\text{OpenList} \gets PriorityQueue()$
            \State $\qquad$ $\text{ClosedList} \gets []$
        \Procedure{A*-Solver($RiskEq$ , $initial$, $B$)}{}
            \State $OpenList.add(initial)$
            \While {$!OpenList.isEmpty()$}
                \State $plan \gets OpenList.poll()$
                \If {$plan \in ClosedList$}
                    \Continue
                \EndIf
                \If {$Cost(plan) \leq B$ }
                        \State \Return $plan$
                \EndIf
                \State $ClosedList.append(plan)$
                \ForEach {$cm \in plan$}
                    \State $newPlan \gets plan \setminus \{cm\}$
                    \If {$!(newPlan \in OpenList) \&\& !(newPlan \in ClosedList)$ }
                        \State $OpenList.add(newPlan)$
                    \EndIf
                \EndFor
            \EndWhile
        \EndProcedure
    \end{algorithmic}
\end{algorithm}

Note that A* can advance in multiple paths simultaneously, while guided by the f-values towards the best solution. 
In our context, the f-value of a node represents the estimated risk in the system under the deployment of a subseset of the plan represented by the node, which cost is equal or lower than the given budget.

To help further guide the search towards a cost-effective solution, the open list is sorted as follows:

\begin{enumerate}
    \item by the f-value -- the lower the better; the plan that can be reduced while introducing as little risk as possible is preferred.
    \item by the g-value -- the higher the better; if the estimate risk introduced by a subset of two plans is similar, the plan that already introduces the greater risk is preferred (this implies that the risk introduced by removing countermeasures is smaller, thus closer to a goal).
    \item by the deployment cost -- the lower the better; if two plans have the exact same risk estimation, the cheaper plan is preferred, since it is closer the budget.
\end{enumerate}

The f-value is calculated based on the current risk in the system (g-value) and the estimated risk to be introduces to the system by removing countermeasures from the current plan towards the budget limitation (h-value). The h-value is computed in two steps as follows:

    \noindent \textbf{ (1) Estimate the minimal amount of countermeasures to remove:} given a node $n = \{C_j=cm_{ID}^{pos} |\forall cm_{ID} \in CM\}$, sort in descending order the deployment costs of the participating countermeasures: $costs_n=\{Cost(C_j) |\forall C_j \in n\}$, and count the top countermeasures (denoted by $X$) required to remove in order to reduce the current plan’s cost to the budget.
    
    \noindent \textbf{(2) Estimate the minimal risk introduced by removing $X$ countermeasures:} given the same node $n$, we define: $n^{-C_j}=n\setminus\{C_j\}$, which is essentially a version of the plan $n$ after removing the countermeasure $C_j$. 
    Then, for each $C_j\in n$ we compute the risk introduced by removing each of the countermeasures in $n$: $risks_n=\{\Delta_n^{C_j}=Risk(n^{-C_j})-Risk(n)|\forall C_j \in n\}$, and sort them in ascending order. 
    Finally, the minimal risk estimated to be added to the system is the sum of the top $X$ values in $risks_n$.

We demonstrated the heuristic calculation using the attack graph in Figure~\ref{fig:mulval-example-ag}. 
Consider the plan $n = \{C_3,C_4,C_5\}$, and a maximal deployment budget of \$50.
First, the costs of $n$'s countermeasures (as specified in Table \ref{tab:example-cm}) are sorted in descending order: $costs_n=\{50,10,10\}$; the total deployment cost of $n$ is \$70. 
If we were to remove the countermeasure that costs \$50 (i.e., $C_4$) from $n$, the new plan's deployment cost will correspond to the budget ($70-50=20 \leq 50$).
Thus, at least one countermeasure should be removed, i.e., $X=1$.

Second, we compute the risk differences ($\Delta_n^{C_j}, j\in \{3,4,5\}$) according to Equation \ref{eq:risk-eq-example}:

\begin{subequations}
    \footnotesize
\[
    \begin{array}{ll}
        n = \{C_3,C_4,C_5\}, & Risk(n) = (0.37 \cdot 0 \cdot 0 \cdot 1 \cdot 1) + (0.65 \cdot 0 \cdot 1 \cdot 1) \\
        n^{-C_3} = \{C_4,C_5\}, & Risk(n^{-C_3}) = (0.37 \cdot 0 \cdot 0 \cdot 1 \cdot 1) + (0.65 \cdot 1 \cdot 1 \cdot 1)  \\
        n^{-C_4} = \{C_3,C_5\}, & Risk(n^{-C_4}) = (0.37 \cdot 1 \cdot 0 \cdot 1 \cdot 1) + (0.65 \cdot 0 \cdot 1 \cdot 1)  \\
        n^{-C_5} = \{C_3,C_4\}, & Risk(n^{-C_5}) = (0.37 \cdot 0 \cdot 1 \cdot 1 \cdot 1) + (0.65 \cdot 0 \cdot 1 \cdot 1)  \\
    \end{array}
\]
    \begin{gather*}
        \Delta_n^{C_3} = 0.65,~~\Delta_n^{C_4} = 0,~~\Delta_n^{C_5} = 0
    \end{gather*}
\end{subequations}

According to the above computation, we get that: $risks_n=\{0,0,0.65\}$. 
Thus the minimal risk for removing one countermeasure from plan $n$ is 0, i.e. $h(n)=0$.

Note that this heuristic is admissible. 
When assessing the risk to be added by a one-countermeasure reduced plan, we essentially inspect the effect of each countermeasure individually, which can be lower or equal to the effect of multiple countermeasures. 
This ensures to not over-estimate the risk introduced. 
Considering the removal of all possible combinations of countermeasures in $X$ is computationally heavy and thus not feasible.

\section{\label{sec:eval}Evaluation}
\begin{figure*}[t]
    \centering
    \includegraphics[scale=0.45]{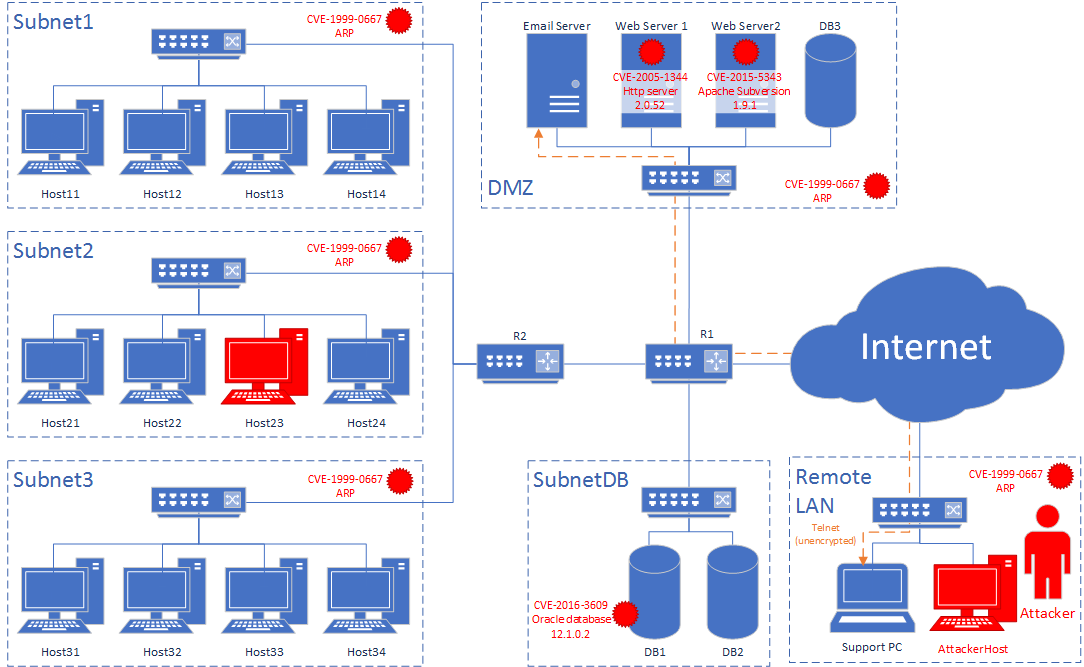}
    \caption{Evaluation environment.}
    \label{fig:eval-env}
\end{figure*}

To demonstrate and evaluate our proposed countermeasure selection process, we modeled the IT network presented in Figure \ref{fig:eval-env}, which contains vulnerabilities that both an external and internal attacker can exploit. We generated an attack graph illustrating all of the different attack paths using MulVAL with our extended modeling, and applied the countermeasure selection process for different budgets.
The heuristic search algorithm we applied is A*
The following subsections describe the modeled network and attacks, and present the results of the selection process for different budgets.

\subsection{Evaluation Environment}
Figure \ref{fig:eval-env} illustrates an organization's IT network which serves as the evaluation environment for the proposed countermeasure selection process. 
The network consists of five subnets in total:

\begin{itemize}
    \item DMZ -- contains two web servers (\textit{Web Server 1} and \textit{Web Server 2}), an e-mail server (\textit{Email Server}) and a database (\textit{DB3}). The DMZ is fully accessible from the Internet.
    \item DB subnet (\textit{SubnetDB}) -- contains the organization's internal databases: \textit{DB1} and \textit{DB2}, which contain sensitive information regarding the organization's clients.
    \item Subnets 1-3 -- contain four hosts each that represents different departments in the organization.
\end{itemize}

Router \textit{R1} connects the DMZ, DB subnet and the internal subnets (via \textit{R2}) with the Internet. 
A vulnerable version of ARP is used across the organization network for intra-subnet communication.
Note that in the current network setup the DMZ is not configured properly since there are no restrictions on external access to the DMZ hosts and there is no separation between the DMZ to the internal subntes.

Employees are allowed to use FTP to transfer data inside the internal network (i.e., among hosts in subnets 1-3), and can access the database servers as part of their work. 
The database servers (\textit{DB1} and \textit{DB2}) run Oracle database software and accept requests on port 1521. 
Note that \textit{DB1} is not patched properly and runs a vulnerable version of the Oracle database software (CVE-2016-3609), which allows data theft.
In addition, \textit{Host11} accepts remote desktop connections from any location in the network.

Remote operators can use an external host (\textit{Support PC}) to connect hosts in the DMZ via Telnet for remote operations. 
The servers in the DMZ also run vulnerable applications: \textit{Web Server 1} runs a vulnerable Apache HTTP server version (CVE-2005-1344) that allows arbitrary code execution upon exploitation; and \textit{Web Server 2} runs a vulnerable Apache Subversion (CVE-2015-5343) that leads to denial of service when exploited.

We consider two types of attackers: internal and external. 
The internal attacker is located in \textit{Host23} that is a part of the internal network (\textit{Subnet2}), while the external attacker is located in the \textit{AttackerHost} that is connected to the Internet.
The external attacker can infiltrate into the network by exploiting the vulnerability in \textit{Web Server 1} that allows code execution and can grant him/her with local access to that server, or by obtaining user credentials to \textit{Email Server} by eavesdropping the Telnet communication between \textit{SupportPC} and \textit{Email Server}.
In both cases the attacker is assumed to gain at least one user account in advance: \textit{host11User}, which allows him/her to open RDP connection to \textit{Host 11}.

To summarize, the attacker's goals are as follows:

\begin{enumerate}
    \item Steal sensitive information from \textit{DB1}.
    \item Spoof the communication between two internal hosts (\textit{Host12} and \textit{Host22}) and see the contents of the data transferred.
    \item Read or send fake e-mails by obtaining local access to \textit{Email Server}.
    \item Tamper with the organization's website by executing  malicious code on \textit{Web Server 1}.
    \item Deny users from accessing the organization's Subversion server (\textit{Web Server 2}) and disrupt their work.
\end{enumerate}

Table \ref{tab:eval-av} describes all of the ways an attacker can gain access to each asset in the network.

\begin{table*}
    \centering
    \caption{Attack vectors in evaluation environment.}
    \scriptsize
    \begin{tabular}{|c|p{0.4\linewidth}|p{0.4\linewidth}|}
        \hline
        \diaghead{\theadfont Attacker Position IIIII}{Target Host}{Attacker Pos.} &  \thead{\textit{Host23} (internal)} &  \thead{\textit{Attacker Host} (external)} \\
         \hline
         
         \textit{Web Server 1} & \makecell[l]{Abuse the free connectivity from the internal subnets to the DMZ} & \makecell[l]{1. Abuse the free connectivity from the Internet to the DMZ \\
         2. Exploit a vulnerability in \textit{Web Server 1} and gain local access} \\
         \hline
         
         \multirow{2}{*}{\textit{Web Server 2}} & Abuse the free connectivity from the internal subnets to the DMZ & Abuse the free connectivity from the Internet \\
         \cline{2-3}
         & \makecell[l]{1. Gain access to \textit{Host11} \\
         2. Abuse the free connectivity from the internal subnets to the DMZ} & \makecell[l]{1. Gain access to \textit{Email Server} or \textit{Web Server 1} \\
         2. Access \textit{Web Server 2} through the LAN} \\
         \hline
         
         \textit{Email Server} & \makecell{--} & \makecell[l]{1. Spoof as \textit{Support PC} \\ 
         2. Abuse the unencrypted Telnet communication to  steal the user's credentials \\
         3. Connect with \textit{Email Server} via Telnet using the stolen credentials} \\
         \hline
         
         \textit{DB1} & \makecell[l]{1. Abuse the free connectivity from the internal subnets to the \textit{SubnetDB} \\ 
         2. Exploit the vulnerability in \textit{DB1}} & 
         \makecell[l]{1. Gain access to \textit{Host11} \\
         2. Abuse the free connectivity from the internal subnets to the \textit{SubnetDB} \\
         3. Exploit the vulnerability in \textit{DB1}} \\
         \hline
         
         \textit{Host11} & \makecell[l]{1. Abuse the free connectivity between the internal subnets \\
         2. Use \textit{host11User} account (obtained in advance) to log in \textit{Host11} via remote desktop} & \makecell[l]{ 1. Gain access to \textit{Web Server 1}, \textit{Web Server 2}, or \textit{Email Server} \\
         2. Use \textit{host11User} account (that was obtained in advance) to log in \textit{Host11}} \\
         \hline
    \end{tabular}
    \label{tab:eval-av}
\end{table*}

\subsection{Results}
To mitigate the risk in the network described above, we consider the countermeasures specified in Table \ref{tab:eval-cm}. 
The costs were assigned in a way that reflects the complexity and the effort required to deploy each type of countermeasure.
There are four cost levels: \$10, \$50, \$300, and \$1000.

\begin{table}[t]
    \centering
    \caption{Countermeasures used for evaluation.}
    \scriptsize
    \begin{tabular}{|c|c|c|c|}
        \hline\multirow{2}{*}{Product} & \multirow{2}{*}{Type} & \multicolumn{2}{c|}{Cost (\$)} \\
        \cline{3-4}
         & & Deploy & Update \\
         \hline
         Cisco Next-Gen Firewall & Network-based firewall & 1000 & 10 \\
         \hline
         ZoneAlarm & Host-based firewall & 300 & 10 \\
         \hline
         Snort & Network-based IPS & 1000 & 10 \\
         \hline
         Wazuh & Host-based IPS & 300 & 10 \\
         \hline
         McAfee Antivirus & \multirow{2}{*}{Antivirus} & \multirow{2}{*}{50} & \multirow{2}{*}{--} \\
         \cline{1-1}
         Kaspersky Antivirus & & & \\
         \hline
         Various & Patch & -- & 10 \\
         \hline
    \end{tabular}
    \label{tab:eval-cm}
\end{table}

Note that when searching for a plan, the algorithm considers only countermeasures for which the cost stands within the budget limitation; e.g., for budget of \$100 the algorithm will not consider host-based IPS as an option, since its cost (\$300) is higher than the specified budget. 
This means that when evaluating a budget that corresponds to one of the cost levels, the search space increases.
Moreover, legitimate services (i.e., the HTTP server in \textit{Web Server 1}, the Apache Subversion in \textit{Web Server 1}, FTP for data transfer, and the communication with the database servers) are crucial to the organization's operation, thus cannot be blocked; Telnet and RDP on the other hand, are less critical, thus can be restricted.

We analyze separately the results for external and internal attack scenarios. 
The best way to secure the network, for both external and internal threats, is: (1) patch the vulnerabilities in \textit{Web Server 1}, \textit{Web Server 2}, and \textit{DB1}; (2) install a network-based IPS on router \textit{R2}; and (3) install a network-based firewall to restrict the RDP communication between the internal subnets and the DMZ and the Telnet communication between external devices and the DMZ hosts (or replace with more protected protocol such as SSH, however this countermeasure is currently not modeled).
The deployment cost of this plan is \$2030.

\begin{table*}
    \centering
    \caption{Recommended plans for each budget.}
    \scriptsize
    \renewcommand{\arraystretch}{1.4}
    \begin{tabular}{c|M{0.4\linewidth}|M{0.4\linewidth}}
         Budget (\$) & Internal Attacker (cost \$) & External Attacker (cost \$) \\
         \hline
         10 & \multicolumn{2}{c}{$Patch^{WebServer1}$ (10)} \\
         \hline
         20 & $Patch^{WebServer1}$, $Patch^{DB1}$ (20) & $Patch^{WebServer1}$, $Patch^{WebServer2}$  (20) \\
         \hline
         30 & \multirow{3}{=}{$Patch^{WebServer1}$, $Patch^{DB1}$, $Patch^{WebServer2}$ (30)} & \multirow{3}{=}{$Patch^{WebServer1}$, $Patch^{WebServer2}$, $Patch^{DB1}$ (30)} \\
         \cline{1-1}
         40 & & \\
         \cline{1-1}
         50 & & \\
         \hline
         100 & \multicolumn{2}{c}{$Patch^{WebServer1}$, $Patch^{DB1}$, $Patch^{WebServer2}$, $AV_2^{WebServer1}$ (80)}  \\
         \hline
         200 & \multicolumn{2}{c}{$Patch^{WebServer1}$, $Patch^{DB1}$, $Patch^{WebServer2}$, $AV_2^{WebServer1}, AV_1^{WebServer1}$  (130)}  \\
         \hline
         300 & \multicolumn{2}{c}{$Patch^{WebServer1}$, $Patch^{DB1}$, $Patch^{WebServer2}$  (30)}  \\
         \hline
         400 & $Patch^{WebServer1}$, $Patch^{DB1}$, $Patch^{WebServer2}$,  $AV_2^{WebServer1}$, $HB-FW^{Host11}$  (380) & $Patch^{WebServer1}$, $Patch^{DB1}$, $Patch^{WebServer2}$, $AV_2^{WebServer1}$, $HB-IPS^{EmailServer}$ (380) \\
         \hline
         500 & \multirow{3}{=}{$Patch^{WebServer1}$, $Patch^{DB1}$, $Patch^{WebServer2}$,  $AV_2^{WebServer1}$, $HB-FW^{Host11}$, $AV_1^{WebServer1}$  (430)} & \multirow{3}{=}{$Patch^{WebServer1}$, $Patch^{DB1}$, $Patch^{WebServer2}$, $AV_2^{WebServer1}$, $HB-IPS^{EmailServer}$, $AV_1^{WebServer1}$ (430)} \\
         \cline{1-1}
         600 &  &  \\
         \cline{1-1}
         700 &  &  \\
         \hline
         800 & \multirow{10}{=}{$Patch^{WebServer1}$, $Patch^{DB1}$, $Patch^{WebServer2}$,  $AV_2^{WebServer1}$, $HB-FW^{Host11}$, $AV_1^{WebServer1}$, $HB-FW^{WebServer1}$ (730)} & \multirow{3}{=}{$Patch^{WebServer1}$, $Patch^{DB1}$, $Patch^{WebServer2}$, $AV_2^{WebServer1}$, $HB-IPS^{EmailServer}$, $AV_1^{WebServer1}$, $HB-FW^{SupportPC}$ (730)} \\
         \cline{1-1}
         900 & & \\
         \cline{1-1}
         1000 & & \\
         \cline{1-1}\cline{3-3}
         1100 &  & \multirow{3}{=}{$Patch^{WebServer1}$, $Patch^{DB1}$, $Patch^{WebServer2}$, $AV_2^{WebServer1}$, $HB-IPS^{EmailServer}$, $AV_1^{WebServer1}$, $HB-FW^{SupportPC}$, $HB-FW^{Host11}$ (1030)} \\
         \cline{1-1}
         1200 &  & \\
         \cline{1-1}
         1300 & & \\
         \cline{1-1} \cline{3-3}
         1400 & & \multirow{3}{=}{$Patch^{WebServer1}$, $Patch^{DB1}$, $Patch^{WebServer2}$, $AV_2^{WebServer1}$, $HB-IPS^{EmailServer}$, $AV_1^{WebServer1}$, $HB-FW^{SupportPC}$, $HB-FW^{Host11}$, $HB-FW^{WebServer1}$ (1330)} \\
         \cline{1-1}
         1500 & & \\
         \cline{1-1}
         1600 & & \\
         \cline{1-1} \cline{3-3}
         1700 & & \multirow{5}{=}{$Patch^{WebServer1}$, $Patch^{DB1}$, $Patch^{WebServer2}$,  $AV_2^{WebServer1}$, $HB-IPS^{EmailServer}$, $AV_1^{WebServer1}$, $HB-FW^{SupportPC}$, $HB-FW^{Host11}$, $HB-FW^{WebServer1}$ $HB-FW^{EmailServer}$ (1630)} \\
         \cline{1-2} 
         1800 & \multirow{4}{=}{$Patch^{WebServer1}$, $Patch^{DB1}$, $Patch^{WebServer2}$, $AV_2^{WebServer1}$, $HB-FW^{Host11}$, $AV_1^{WebServer1}$, $HB-FW^{WebServer1}$, $NB-FW^{Subnet2-Subnet1}$ (1730)} & \\
         \cline{1-1}
         1900 & & \\
         \cline{1-1}
         2000 & & \\
         \cline{1-1}
         2100 &  & \\
         \hline
    \end{tabular}
    \label{tab:plans-per-budget}
\end{table*}

\begin{figure}
    \centering
    \includegraphics[scale=0.55]{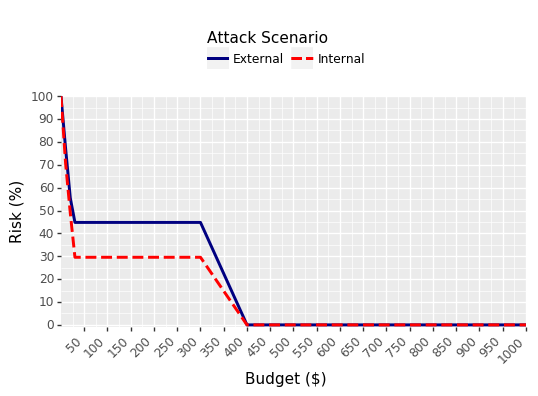}
    \caption{Risk per budget.}
    \label{fig:eval-risk-budget}
\end{figure}

We examined the countermeasure plans generated for different budgets ranging from \$10 to \$2100.
The plans for each tested budget are presented in Table \ref{tab:plans-per-budget}.
Figure \ref{fig:eval-risk-budget} presents the risk level under the deployment of each generated plan for each attack scenario. 
The plot presents the risk for the plans suggested only for budgets \$50-\$1000.
Note that the plans for budgets \$1100-\$2100 also eliminate completely the risk in the system, thus not included in the plot.

The suggested plans demonstrate the great advantage of using an attack graph-based risk assessment.
The attack graph specifies the actions that the attacker should take from his/her initial position to his/her goals. 
This enables to capture stepping stones that are common to multiple attack paths and to understand which countermeasure can eliminate attack paths as soon as possible.
One can see that as the budget increases, the algorithm chooses the countermeasure not only by the amount of risk they mitigate and cost, but also according to the attack steps; i.e., the algorithm strives to block the attack as early as possible.
For example, consider budgets \$10-\$20, which permit patching/updating one or two devices. 
The plan for budget \$10, in both scenarios, includes patching only \textit{Web Server 1} since it introduces the highest amount of risk. 
However, the plans for budget \$20 differ.For the external attacker, the DMZ servers (\textit{Web Server 1}, \textit{Web Server 2}, and \textit{Email Server}) are the stepping stones in his/her way towards the internal network; thus, in this scenario, the risk they introduce is higher compared to the vulnerabilities in the external and the algorithm suggest to further patch \textit{Web Server 2}. 
On the other hand, in the internal attack scenario, the attacker is more accessible to the internal hosts, thus the algorithm suggest to further patch \textit{DB1}.
This is also evident in the plans suggested for budget \$400. 
For both scenarios, patching \textit{Web Server 1}, \textit{Web Server 2}, and \textit{DB1} eliminate most of the attack vectors. 
In the internal attack scenario, the only attack goal that survives the patching is spoofing the communication between \textit{Host12} and \textit{Host 22}, since the attacker can still log in to \textit{Host 11} (using RDP) and execute the attack; while in the external attack scenario, the attacker can still exploit the vulnerable communication with the \textit{Email Server} to compromise it, and also to further execute the spoofing attack. As can be seen in the results, the algorithm succeed to capture this information and suggested to extend the plan with a host-based firewall on \textit{Host11} to restrict RDP connections and host-based IPS on \textit{Email Server} accordingly.

Note that as the budget increases, the plans also become redundant (i.e., contain several countermeasures that tackle the same vulnerability).
These redundancies suggest to protect the system's assets with multiple layers of protection.

For example, besides patching a specific vulnerability in \textit{Web Server 1}, the code execution can be detected by antivirus software; thus, starting from budget of \$100, installing antivirus software on \textit{Web Server 1} is included in the suggested plans together with patching it. 
Another example is restricting the incoming/outgoing communication to/from hosts that expose vulnerable services. 
The suggested plans for budget \$800 include host-based firewalls.
In the internal scenario, a firewall for \textit{Web Server 1} is suggested to restrict its RDP communication with \textit{Host11}, in addition to suggesting to install firewall on \textit{Host11} to restrict incoming RDP connections.
In the external scenario, it is suggested to restrict the outgoing Telnet communication from the \textit{Support PC} by a host-based firewall on top of installing host-based IPS on the \textit{Email Server} to identify suspicious logins.

Another interesting result is the plan suggested for budget \$300. 
In the internal scenario, the only attack goal that survives the patching is spoofing the communication between \textit{Host12} and \textit{Host 22}, since the attacker can still log in to \textit{Host 11} (using RDP) and execute the attack. 
To thwart this attack, a host-based firewall can be installed on \textit{Host 11} to prevent RDP connections. 
On the other hand, in the external scenario, the attacker's access to the internal subnets depends solely on exploiting the vulnerabilities in the DMZ servers.
At this point, only \textit{Email Server} is not patched and can grant the attacker access to the internal network. 
This attack vector can be eliminated by installing a host-based IPS on \textit{Email Server}. 
As mentioned above, host-based firewall on \textit{Host11} (for the internal scenario) and host-based IPS on \textit{Email Server} (for the external scenario) are the two countermeasures that can eliminate all of the remaining risk in the system after patching all possible vulnerabilities (i.e., the plan for budget \$30). 
However, the three patches reduces more risk than just a firewall or IPS, thus the algorithm suggested patching over firewall or IPS. 
The reason that the plan doesn't include antivirus software as in the plans for \$100-\$200, is the fact that these plans have the same f-value and g-value (because they cover the same vulnerabilities and are within the budget limitation), thus the cheapest plan is preferred (see subsection \ref{subsec:astar}).

The plans for budget \$2100 contain many of the countermeasures we expected to see; however, the algorithm didn't suggest on the expected network-based solutions. 
The plan suggested for the internal scenario includes: all three patches, both antivirus software for \textit{Web Server 1}, host-based firewalls for \textit{Host11} and \textit{WebServer1} to restrict RDP connections to \textit{Host11}, and a network-based firewall between \textit{Subnet2} and \textit{Subnet1} (also to restrict RDP connections to \textit{Host11}.
The plan suggested for the external scenario includes: all three patches, both antivirus software for \textit{Web Server 1}, a host-based IPS for \textit{Email Server}, host-based firewalls for \textit{Support PC} and \textit{Email Server} (to restrict Telnet), \textit{Host11} and \textit{Web Server 1} (to restrict RDP).
The network modeling we used to generate the attack graph considers firewalls to control the communication between only two subnets.
Thus our approach proposes to place a network-based firewall between the Internet and DMZ, and between the DMZ and \textit{Subnet 1}. 
Since there are only three types of communication to block between the Internet and DMZ, the algorithm prefers to place three host-based solutions (that cost \$900) over a network-based solution (that costs \$1000).
A network-based firewall solution was nor selected to control the communication between the Internet and DMZ due to its cost. 

Note that currently there is no preference for countermeasure of the same type; e.g., one antivirus software is not preferred over the other, thus the algorithm recommendation on the specific antivirus software is random. 
This can be changed by introducing preference criterion to the countermeasure modeling and integrating it in the risk assessment or during the search.

\begin{figure}
    \centering
    \includegraphics[scale=0.55]{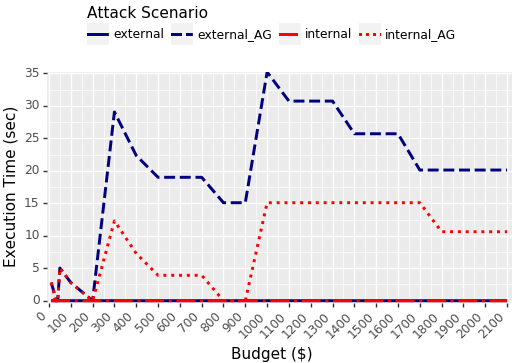}
    \caption{Calculating risk using equations vs. regenerating attack graph.}
    \label{fig:ag_vs_eq_exec_time}
\end{figure}

\subsubsection{Attack graph regeneration vs. risk equations}
Unlike other methods (e.g., \cite{santhanam2013identifying,chung2013nice,kotenko2016dynamic}), we avoid regenerating the attack graph when evaluating the effectiveness of a countermeasure plan, since it is computationally heavy. 
Instead, we suggest to represent the attack graph and the countermeasures that can eliminate each node as mathematical equations, where the variables correspond to the implementation of each possible countermeasure, and just substitute them with suitable values according to a given plan (see Subsection \ref{subsec:risk-eq}).

Figure \ref{fig:ag_vs_eq_exec_time} illustrates the difference in the execution time for calculating the risk when using our approach and regeneration of the attack graph under different plans. 
We measured the attack graph regeneration time by executing the MulVAL tool (as we do in the risk assessment phase) with various input files simulating the application of different countermeasure plans, and took the minimal time as reference.
This means that computing the risk for \textbf{each} expanded node during the search will take at least this minimal time measurement.
Note the this measurement is an under estimation of the real attack graph generation time. 
As can be observed, regenerating the attack graph to compute the risk is significantly less efficient than our suggested approach.

\subsubsection{Attack graph-based vs. traditional risk assessment}
Traditional risk assessment process, such as performed in \cite{viduto2012novel}, is mostly manual. 
The first step is to gather information about the network and its vulnerabilities. 
This information is crossover with public data sources, e.g., NVD and CVSS.
Next, a threat analysis is performed. 
This step includes identifying potential threat sources and associate the discovered vulnerabilities to threats. The likelihood of a risk acting over a vulnerability is estimated based on the organization's history.
Finally, a list of generic countermeasures is generated. Each countermeasure has its cost (that consist of purchase cost, operational cost, training cost, and man power) and a score that represents its impact on the vulnerabilities that was found (which is usually the base CVSS score). 
The countermeasures are chosen by the severity scaling of vulnerabilities; i.e., the vulnerabilities with the highest CVSS score are treated first.

The evaluation network contains four vulnerabilities with CVEs: (1) CVE-1999-0667 with CVSS score 10; (2) CVE-2005-1344 with CVSS score 7.5; (3) CVE-2015-5343 with CVSS score 7.6; and (4) CVE-2016-3609 with CVSS score 9. 
Moreover, the Telnet protocol is vulnerable since it is not encrypted. 
This is a vulnerability in the protocol's design, thus doesn't have a specific CVE and required to be scored by an expert. 
Since it allows stealing users' credentials, it is reasonable to be assigned with the maximum score of 10.

If we were to perform a traditional risk assessment to the evaluation network, we would choose to fix the vulnerabilities in the following order: (1) CVE-1999-0667 and Telnet; (2) CVE-2016-3609; (3) CVE-2015-5343; and lastly CVE-2005-1344.
This means that we will first fix the vulnerabilities on the internal hosts and only then the vulnerability in the DMZ host. 
However, when considering an external attacker, many of the attack vector can be eliminated by fixing the DMZ hosts vulnerabilities, which have less severely ranked.

The most severe vulnerability is the ARP spoof (CVE-1999-0667). 
To fix this, the protocol should be replaced, or defense mechanisms should be installed all on hosts in all subnets. 
Fixing this vulnerability is important but also entails high expenses.
However, note that in order for an external attacker to reach the internal network, he/she is required to first gain access to one of the hosts in the DMZ. 
Thus, the attacker can be prevented from reaching the internal network by fixing the vulnerabilities in these hosts, although most of them are less severe than the internal network vulnerabilities. 
Note that fixing these vulnerabilities is also likely to be less expensive, since it requires to patch two hosts and to change or restrict the communication with a specific host.

These insights cannot be reached without considering the network topology, thus choosing countermeasures based just on vulnerabilities' severity doesn't guarantee an optimal plan. 
This is exactly the missing information that is captured by attack graphs.
Since attack graphs illustrate the paths from the attacker's initial position to his/her goal, it captures the fact that fixing the DMZ hosts first will eliminate the paths to the internal hosts (this step will appear at the beginning of the path); and this is reflected in the plans that our algorithm provides.

\section{\label{sec:conclusion}Conclusion and Future Work}
We present a heuristic approach towards the security countermeasure selection problem.
Our approach uses an attack graph-based risk assessment process that enables countermeasure selection that consider the network topology, the vulnerabilities found, and the interactions between them; thus provides more accurate and applicable suggestions. 
Moreover, we provide an automated method for identifying which countermeasure can eliminate each step in each attack path.
Our approach also explicitly specifies the network position where each countermeasure should be applied.

We model the countermeasure selection problem as a graph path finding problem and used a heuristic solver (A*) to find an optimal solution: a plan (i.e., set of countermeasures and their deployment position) that minimizes the risk in the system as much as possible within the given budget limitation.
The results show that our approach was able to provide solution very close to our expectations. 
Moreover, since we use an attack graph-based risk assessment, the plans suggested strive to block each attack path as soon as possible (i.e., in its initial stages). 
The proposed plans are also redundant when the budget permits, and may provide suggestions for multiple countermeasures that tackle the same vulnerability or attack path to enhance the security of the system.

Currently, our approach filters out countermeasures that contradict the organization policy (i.e, not blocking ports of legitimate services). 
In future work, we aim to extend it to consider operators' preferences. 
This can be done as part of the relevant countermeasure identification phase (e.g., in case that the operator doesn't want to deploy a specific countermeasure on a specific host), or by introducing another cost component to make one countermeasure preferable over the other during the search.
Another aspect that should taken into account is shared costs (e.g., one antivirus licence can cover multiple hosts), which significantly affects the algorithm's decisions.

\bibliographystyle{ACM-Reference-Format}
\bibliography{main.bib}

\appendix

\section{\label{app:ma-def-example}Mitigation Actions - Example}
    For demonstrating the definition of mitigation actions corresponding to the \textit{vulHost} and \textit{aclNW} predicates, consider the predicate and rules in Listing \ref{listing:vilhost_aclnw}.
\\
\textbf{Step 1: create defense mechanism set.}  As specified in Subsection \ref{subsec:ma}, we consider the following: host and network based firewalls, host and network based IPS, software patch, and antivirus software, i.e.:
\[
DM=\{HB-FW, NB-FW, HB-IPS, NB-IPS, Patch, Antivirus\}
\] \\
\textbf{Step 2: match defense mechanisms to predicates.} Host vulnerabilities (\textit{vulHost}) can be fixed by patching the vulnerable software or by capturing and blocking the abnormal behavior indicating on that vulnerability by a network-based IPS, thus:
\[ DM_{vulHost}=\{Patch, NB-IPS\} \]
Connectivity among hosts across the network (\textit{aclNW}) can be controlled by network-based firewalls, thus:
\[ DM_{aclNW}=\{NB-FW\} \] \\
\textbf{Step 3: represent the scenarios to be cancelled.} \textit{vulHost} is a primitive predicate and thus: 
\[ p_{vulHost}^{pre}=\phi \]
\[ p_{vulHost}^{post}=\{vulHost(Host, VulID, Prog, Range, Consequense)\} \]
\[ S_{vulHost}=\{(p_{vulHost}^{post}, p_{vulHost}^{pre}\}) \] \\
On the other hand, \textit{aclNW} is a multi-rule, thus we first have to group together the rules that represent similar scenarios:
\[ g_{aclNW}^1 = \{r^1_{aclNW},r^2_{aclNW}\} \]
\[ g_{aclNW}^2 = \{r^3_{aclNW}\} \]
\[ g_{aclNW}^3 = \{r^4_{aclNW}\} \]
\[ g_{aclNW}^4 = \{r^5_{aclNW}\} \] \\
Group $g^1_{aclNW}$ represents rules for host to host connectivity, $g^2_{aclNW}$ represents rules for subnet to host, $g^3_{aclNW}$ represents rules for host to subnet, and $g^4_{aclNW}$ represents the rule that indicate on subnet to subnet. 
The pre-conditions of each group are as follows:
\[ 
\begin{split}
    (g_{aclNW}^1)^{pre} =& \{ located(SrcHost, SrcSubnet, ipSubnet), \\
                        & located(DstHost, DstSubnet, ipSubnet)\}
\end{split} 
\]

\[ 
\begin{split}
    (g_{aclNW}^2)^{pre} = \{ located(DstHost, DstSubnet, ipSubnet)\}
\end{split} 
\]

\[ 
\begin{split}
    (g_{aclNW}^3)^{pre} = \{ located(SrcHost, SrcSubnet, ipSubnet)\}
\end{split} 
\]

\[ 
\begin{split}
    (g_{aclNW}^4)^{pre} = \phi
\end{split} 
\]
Note that there are no pre-conditions for the fourth group since it is essentially the primitive representation of this predicate, thus it is handled as in the first case. Finally, the scenarios for \textit{aclNW} are:
\[
    \begin{split}
        S&_{aclNW} = \{(aclNW(SrcHost, DstHost, Protocol, Port),(g_{aclNW}^1)^{pre}), \\
                    &(aclNW(SrcSubnetOrHost, DstHost, Protocol, Port),(g_{aclNW}^2)^{pre}),\\
                    &(aclNW(SrcHost, DstHostOrSubnet, Protocol, Port), (g_{aclNW}^3)^{pre}),\\
                    &(aclNW(SrcHostOrSubnet, DstHostOrSubnet, Protocol, Port), (g_{aclNW}^4)^{pre}) \}
    \end{split}
\] \\
\textbf{Step 4: create mitigation action instances.} As specified in step 2, the possible mitigation actions are: patching, NIPS, and network-based firewall. In order to deploy an IPS or patch a program there must exist a suitable rule or patch. 
Thus the following are the pre-conditions for deploying IPS and install a patch correspondingly:
\[ ipsRule(VulID, RuleID) \]
\[ hasPatch(Program, VulID, PatchID) \]
Moreover, IPS and firewall are mechanisms that might require installation, or just an update (if already installed). 
We distinguish between these cases by indicating in the pre-conditions on the existence of such components using the following predicates:
\[ isFirewall(FW, SrchSubnet, DstSubnet) \]
\[ isIPS(IPS, Subnet) \]
NIPS however, requires additional pre-conditions to be deployed, such that the host it is installed in should be a gateway of the subnet it monitors; this is represented by the following predicate:
\[ isGateway(Host,Subnet) \]

Table \ref{tab:ma-def-example} presents all of the mitigation actions instances created in this process. 
Mitigation actions 1-4 are relevant to the \textit{vulHost} predicate, while 5-12 to \textit{aclNW} predicate. More specifically, actions 5-6 represent $g_{aclNW}^1$, 7-8 represent $g_{aclNW}^2$, 9-10 represent $g_{aclNW}^3$, and 11-12 represent $g_{aclNW}^4$.

\lstinputlisting[
  style      = Prolog-pygsty,
  caption    = {Example \textit{vulHost} and \textit{aclNW} predicates.},
  label = {listing:vilhost_aclnw},
  float=tp,
  floatplacement=tbp
]{rules.pl}

\onecolumn
\begin{table*}[th!]
    \centering
    \caption{Mitigation actions for \textit{vulHost} and \textit{aclNW} predicates.}
    \scriptsize
    \renewcommand{\arraystretch}{1.2}
    \begin{tabular}{|c|c|m{0.25\textwidth}|c|c|}
        \hline
        ID & Type & \centering Description & Pre-conditions & Post-conditions \\
        \hline
        1 &  Patch & Patch the vulnerable program & $hasPatch(Program,VulID,PatchID)$ & $vulHost(Host,VulID,Program,Range,Consequence)$ \\
        \hline
        \multirow{4}{*}{2} & \multirow{4}{*}{NB-IPS} & \multirow{4}{=}{Place new IPS in the gateway} & $ipsRule(VulID,RuleID)$ & \multirow{2}{*}{$vulHost(Host,VulID,Program,Range,Consequence)$} \\
        \cline{4-4}
        & & & $!isIPS(IPS,Subnet)$ & \\
        \cline{4-5}
        & & & $isGateway(IPS,SubnetA)$ & \multirow{2}{*}{$isIPS(IPS,SubnetA)$}\\
        \cline{4-4}
        & & & $located(Host,SubnetA,ipSubnet)$ & \\
        \hline
        \multirow{3}{*}{3} & \multirow{3}{*}{NB-IPS} & \multirow{3}{=}{Update rules of an existing IPS} & $ipsRule(VulID,RuleID)$ & \multirow{3}{*}{$vulHost(Host,VulID,Program,Range,Consequence)$} \\
        \cline{4-4}
        & & & $isIPS(IPS,Subnet)$ & \\
        \cline{4-4}
        & & & $located(Host,Subnet,ipSubnet)$ & \\
        \hline
        \multirow{5}{*}{4} & \multirow{5}{*}{NB-IPS} & \multirow{5}{=}{Connect subnet to existing IPS} & $ipsRule(VulID,RuleID)$ & \multirow{2}{*}{$vulHost(Host,VulID,Program,Range,Consequence)$}\\
        \cline{4-4}
        & & & $isIPS(IPS,Subnet)$ & \\
        \cline{4-5}
        & & & $isGateway(IPS,SubnetA)$ & \multirow{3}{*}{$isIPS(IPS,SubnetA)$} \\
        \cline{4-4}
        & & & $!isIPS(IPS,SubnetA)$ & \\
        \cline{4-4}
        & & & $located(Host,SubnetA,ipSubnet)$ & \\
        \hline
        \multirow{3}{*}{5} & \multirow{3}{*}{NB-FW} &  \multirow{3}{=}{Place new firewall between the subnets of \textit{SrcHost} and \textit{DstHost}} & $!isFirewall(FW,SrcSubnet,DstSubnet)$ &  $aclNW(SrcHost,DstHost,Protocol,Port)$ \\
        \cline{4-5}
        & & & $located(SrcHost,SrcSubnet,ipSubnet)$ &  \multirow{2}{*}{$isFirewall(FW,SrcSubnet,DstSubnet)$} \\
        \cline{4-4}
        & & & $located(DstHost,DstSubnet)$ & \\
        \hline
        \multirow{3}{*}{6} & \multirow{3}{*}{NB-FW} &  \multirow{3}{=}{Update existing firewall between the subnets of \textit{SrcHost} and \textit{DstHost}} & $isFirewall(FW,SrcSubnet,DstSubnet)$ &  \multirow{3}{*}{$aclNW(SrcHost,DstHost,Protocol,Port)$} \\
        \cline{4-4}
        & & & $located(SrcHost,SrcSubnet,ipSubnet)$ &  \\
        \cline{4-4}
        & & & $located(DstHost,DstSubnet)$ & \\
        \hline
        \multirow{2}{*}{7} & \multirow{2}{*}{NB-FW} &  \multirow{2}{=}{Place new firewall between \textit{SrcSubnet} and \textit{DstHost}'s subnet} & $!isFirewall(FW,SrcSubnet,DstSubnet)$ &  $aclNW(SrcSubnet,DstHost,Protocol,Port)$ \\
        \cline{4-5}
        & & & $located(DstHost,DstSubnet,ipSubnet)$ & $isFirewall(FW,SrcSubnet,DstSubnet)$ \\
        \hline
        \multirow{2}{*}{8} & \multirow{2}{*}{NB-FW} &  \multirow{2}{=}{Update existing firewall between \textit{SrcSubnet} and \textit{DstHost}'s subnet} & $isFirewall(FW,SrcSubnet,DstSubnet)$ &  \multirow{2}{*}{$aclNW(SrcSubnet,DstHost,Protocol,Port)$} \\
        \cline{4-4}
        & & & $located(DstHost,DstSubnet,ipSubnet)$ & \\
        \hline
        \multirow{2}{*}{9} & \multirow{2}{*}{NB-FW} &  \multirow{2}{=}{Place new firewall between \textit{DstSubnet} and \textit{SrcHost}'s subnet} & $!isFirewall(FW,SrcSubnet,DstSubnet)$ &  $aclNW(SrcHost,DstSubnet,Protocol,Port)$ \\
        \cline{4-5}
        & & & $located(SrcHost,SrcSubnet,ipSubnet)$ & $isFirewall(FW,SrcSubnet,DstSubnet)$ \\
        \hline
        \multirow{2}{*}{10} & \multirow{2}{*}{NB-FW} &  \multirow{2}{=}{Update existing firewall between \textit{DstSubnet} and \textit{SrcHost}'s subnet} & $isFirewall(FW,SrcSubnet,DstSubnet)$ &  \multirow{2}{*}{$aclNW(SrcHost,DstSubnet,Protocol,Port)$} \\
        \cline{4-4}
        & & & $located(SrcHost,SrcSubnet,ipSubnet)$ & \\
        \hline
        \multirow{2}{*}{11} & \multirow{2}{*}{NB-FW} &  \multirow{2}{=}{Place new firewall between \textit{SrcSubnet} and \textit{DstSubnet}} & \multirow{2}{*}{$!isFirewall(FW,SrcSubnet,DstSubnet)$} &  $aclNW(SrcSubnet,DstSubnet,Protocol,Port)$ \\
        \cline{5-5}
        & & & & $isFirewall(FW,SrcSubnet,DstSubnet)$ \\
        \hline
        12 & NB-FW &  Update existing firewall between \textit{SrcSubnet} and \textit{DstSubnet} & $isFirewall(FW,SrcSubnet,DstSubnet)$ &  {$aclNW(SrcHost,DstSubnet,Protocol,Port)$} \\
        \hline
    \end{tabular}
    \label{tab:ma-def-example}
\end{table*}

\end{document}